\documentclass[12pt]{article}
\usepackage{bm,amssymb}
\usepackage{hyperref}
\usepackage[pdftex]{graphicx}

\newtheorem{theorem}{Theorem}
\newcommand{\D}{{\cal D}}
\newcommand{\J}{{\cal J}}
\newcommand{\bbr}{{\bf r}}
\newcommand{\bv}{{\bf v}}
\newcommand{\ba}{{\bf a}}
\newcommand{\p}{\Phi}
\newcommand{\q}{\Psi}
\newcommand{\Tr}{{\rm Tr}}

\begin{document}
\title{Control landscape for ultrafast manipulation by a qubit}
\author{Alexander Pechen$^{1}$\footnote{E-mail: \href{apechen@gmail.com}{apechen@gmail.com}; Webpage: \href{http://www.mathnet.ru/eng/person17991}{mathnet.ru/eng/person17991}} \, and Nikolay Il'in$^{2}\footnote{E-mail: \href{ilyn@mi.ras.ru}{ilyn@mi.ras.ru}}$}
\date{}
\maketitle
\vspace{-1cm}
\begin{center}
$^{1}$Steklov Mathematical Institute of Russian Academy of Sciences\\ Gubkina str., 8, Moscow 119991, Russia\\
$^{2}$The National University of Science and Technology MISiS\\
Leninsky Prospekt, 4, Moscow 119049, Russia
\end{center}

\begin{abstract}
In this work we study extrema of objective functionals for ultrafast manipulation by a qubit. Traps are extrema of the objective functionals which are optimal for manipulation by quantum systems only locally but not globally. Much effort in prior works was devoted to the analysis of traps for quantum systems controlled by long enough laser pulses and, for example, manipulation by a qubit with long control pulses was shown to be trap-free.  Ultrafast femtosecond and attosecond control becomes now widely applicable that motivates the necessity for the analysis of traps on the ultrafast time scale. We do such analysis for a qubit and show that  ultrafast state transfer in a qubit remains trap-free for a wide range of the initial and final states of the qubit. We prove that for this range the probability of transition between the initial and the final states has a saddle but has no traps.
\end{abstract}

\section{Introduction}
The control of quantum  atomic and molecular systems is an important branch of modern science with multiple existing and prospective applications in physics, chemistry, and quantum technologies~\cite{BrumerBook,RiceBook,TannorBook,LetokhovBook,AlessandroBook,Brif2012,Lidar2010,Glaser2015Report}. Ultrafast femtosecond and attosecond laser pulses become  prospective tool~\cite{Brumer1999} which is used for coherent control of two-photon transitions~\cite{Meshulach1998}, laser control of charge transport~\cite{Franco2008,Falke2014}, production of molecular excitations to high rotational states with a preferred sense of rotation~\cite{Prior2008,Prior2012}, tracking the real-time motion of electrons in ions and molecules~\cite{Smirnova2010,Calegari2014}, control of molecular dynamics~\cite{Gerber2007}, creation of quantum coherences in single organic molecules~\cite{Hilder2011}, bond making~\cite{Levin2015-1,Levin2015-2}, etc. Various algorithms were proposed and applied for controlling quantum systems
including gradient ascent pulse engineering (GRAPE)~\cite{GRAPE}, Krotov-type
methods~\cite{Tannor1992}, machine learning~\cite{Sanders2011}, the Broyden-Fletcher-Goldfarb-Shanno (BGFS) algorithm and its
modifications, genetic algorithms and evolutionary strategies~\cite{Judson2002}, quantum annealing\cite{Lidar2014} and combined approaches~\cite{Eitan2011,Goerz2015}. Gradient flows are widely used for optimization in quantum information and quantum dynamics~\cite{Glaser2010}.  Maximum speed limit allowed by quantum evolution was analyzed~\cite{Caneva2009,Lidar2015} as well as stabilization and convergence speed for Hamiltonian control of quantum dynamical semigroups~\cite{Ticozzi2012}. Time optimal control was studied for spin systems~\cite{Khaneja2001} as well as for state preparation with dissipative dynamics~\cite{Lapert2010} and generation of $SU(2)$ operations~\cite{Garon2013} in a single two-level system.

On ultrafast time scale the influence of the environment on the system is often negligible and system's evolution under the action of coherent control $f(t)$, e.g., a shaped laser pulse, can be approximately described by the unitary evolution operator $U_t^f$ which satisfies the time dependent Schr\"odinger equation
\begin{equation}\label{e00}
i\frac{dU^f_t}{dt}=(H_0+f(t)V)U^f_t,\qquad U^f_{t=0}=\mathbb I.
\end{equation}
Here $H_0$ is the free system Hamiltonian and $V$ is the  Hamiltonian for interaction of the system with the laser pulse. For an $n$-level quantum system, $H_0$ and $V$ are $n\times n$ Hermitian matrices.

A particular important class of  control problems can be formulated as maximization of an objective functional of the form
\begin{equation}
{\cal J}_A[f] = {\rm Tr}(\rho_T A)={\rm Tr}(U^f_T\rho_0U^{f\dagger}_T A)
\end{equation}
where $\rho_T=U^f_T\rho_0U^{f\dagger}_T$ is the system density matrix at some final time $T$, $\rho_0$ is the system density matrix at the initial time $t=0$ and $A$ is a Hermitian matrix. The matrix $A$ describes a quantum mechanical observable of the system, e.g., its energy, population of some state, etc. The objective ${\cal J}_A[f]$ describes average value of  $A$ at time $T$ when the system evolves under the action of the control $f$.  A special case of the objective functional  ${\cal J}_A[f]$ is the functional ${\cal J}_{P_{\psi}}[f]$ which describes the probability of finding the system in the state $|\psi\rangle$, ${\cal J}_{P_{\psi}}[f]=|\langle\psi|U^f_T |\psi\rangle|^2$. The corresponding observable  $A=P_{\psi}=|\psi\rangle\langle\psi|$ is the projector on the vector  $|\psi\rangle$.  The control goal is to find a function $f(t)$ that maximizes the  objective ${\cal J}_A[f]$.

The analysis of extrema of the objective functionals is an important problem in quantum control theory. High interest is directed towards the problem of existence or non-existence of locally but not globally optimal controls, the so-called traps~\cite{Rabitz2004,Rabitz2006-1,Rabitz2006-2,RabitzHo,PechenTannorPRL2012,Pechen,Schirmer2013,Mian,Calarco2015}. Extrema of the objective functionals for control of open quantum systems were studied in~\cite{Wu2008,Pechen2008} and the results were applied to build OptiChem theory for optimization in quantum physics and chemistry~\cite{Moore2011-1,Moore2011-2}. Any small variations of a trapping control do not improve the objective. Thus locally traps look as optimal controls, while globally they can be far from optima. A local, e.g., gradient, optimization algorithm can stop the search at a trap thereby not finding a globally optimal solution. If the objective functional has traps with large attraction domain, then chances that the algorithm will reach a solution close to a desired global optimum can become small. Because of this property traps, if they exist, may significntly hinder the search for globally optimal controls and this circumstance motivates the importance of their analysis. Establishing the trap-free  property of a control problem would imply the possibility of easy finding of respective globally optimal controls.

The general absence of traps for common quantum control problems was suggested in~\cite{Rabitz2004,RabitzHo}. Trapping behavior for some systems with number of levels $n\ge 3$ was found in~\cite{PechenTannorPRL2012}, where the presence of second-order traps, i.e. critical points of the objective functional ${\cal J}_A[f]$ which are not global maxima and where the Hessian of ${\cal J}_A[f]$ with respect to $f$ is   negative-semidefinite but not necessarily negative definite, was found. The absence of traps was proved for maximization of transition probability and generation of unitary quantum operations in two-level quantum systems for sufficiently long duration of the control pulses~\cite{Pechen, Mian}. Trap-free behavior was also found for the problem of optimization of the transmission coefficient for a quantum particle passing through potential whose shape is used as control~\cite{PechenTannorCJC2014}. In these works the control time $T$ was assumed to be sufficiently long.

Ultrafast control is more difficult to implement. This difficulty arises, on the one hand, due to the fact that the full controllability of the system can be guaranteed only for a sufficiently long control time $T$. On the other hand,  it is technically difficult to generate ultrashort modulated laser pulses. Another obstacle is associated with quantum speed limit which is the fundamental bound for maximum speed at which a quantum system can evolve in its space of states~\cite{Caneva2009,Lidar2015}. For example, for Landau--Zener system the existence of the time $T_{QSL}$ was demonstrated such that for a pulse duration $T$ smaller than $T_{QSL}$ the Krotov optimization algorithm will not converge while for $T>T_{QSL}$ the value of the fidelity decreases exponentially during iterations~\cite{Caneva2009}. In the context of ultrafast control, an important question is if the control problem will remain trap-free when control time $T$ becomes small, or for small $T$ traps will arise in originally trap-free systems. In this work we analyze this problem for ultrafast manipulation by a single qubit.

According to the analysis of~\cite{Mian}, only single exceptional constant control $f=f_0$ could be a trap for small $T$. All other controls were proven to be not traps for any $T$. In this paper we show that for a wide range of the initial states $\rho_0$,  target observables $A$ and Hamiltonians $H_0, V$ of the qubit, the exceptional  control $f=f_0$ is a saddle point but not a trap for any arbitrary small final time $T$, and hence control problem is trap-free for arbitrary fast control.

\section{Parameters of the control problem}

Define the constant control
\begin{equation}
f_0:=-\frac{\Tr(H_0V)}{\Tr(V^2)}.
\end{equation}
The control $f_0$ will be called {\it exceptional}. This exceptional control plays a crucial role in the analysis below since only this control can potentially be a trap. The control $f_0$ is the only control which if critical (i.e., such that  ${\rm grad} \J_A[f_0]=0$) is also non-regular, i.e., such that the Jacobian of the map $f\to U_T^f$ evaluated at $f_0$ has not full rank. The physical interpretation of this property of the control $f_0$ is the following. Let apply to the system a laser pulse of constant intensity $f_0$. It will produce some unitary evolution  $U_{T}^{f_0}$ of the system. The non-regularity of the pulse means that its arbitrary small modulations can not produce all possible small variations of the unitary evolution  $U_{T}^{f_0}$; there exist such small variations of the evolution operator which can not be produced by any small modulations of $f_0$. If $f_0$ would be a trap for problem of transfer some initial state $|\psi_{\rm i}\rangle$ into some final state $|\psi_{\rm f}\rangle$, it would imply that probability of transition from the initial to the final state under the action of the pulse $f_0$ would be less than maximal while making any small modulations of the pulse would not improve the transition probability.
	
We also define the special time $T_0$
\begin{equation}
T_0:=\frac{\pi}{\|H_0-(1/2){\rm Tr}H_0+f_0V\|}
\end{equation}

The following statement  implies absence of traps for two-level quantum systems if control pulses are sufficiently long~\cite{Mian}.
\begin{theorem}
If ${\rm Tr} V=0$ and $T\geq T_0$, then all maxima and minima of the objective functional~${\cal J}_A[f]={\rm Tr}(U_T^f\rho_0U^{f\dagger}_TA)$ are global. Any control $f\neq f_0$ can not be a trap for any $T>0$.
\end{theorem}
Hence for $T\geq T_0$ there are no traps at all and  for small $T$ only the exceptional control $f=f_0$ could be a trap.

In many situations the interaction Hamiltonian $V$ has zero matrix elements in the basis of eigenvectors of the free Hamiltonian $H_0$, so that in particular, ${\rm Tr}V=0$ and ${\rm Tr}(H_0V)=0$. Under these conditions the Schr\"odinger equation~(\ref{e00}) after suitable change of basis and time rescaling  can be rewritten~\cite{Mian} as
\begin{equation}\label{e01}
i\frac{dU^{f}_t}{dt} =\bigg(\sigma_z+f(t)(v_x\sigma_x+v_y\sigma_y)\bigg)U^{f}_t
\end{equation}
In this case $f_0=0$ and therefore the transition probability for small $T$ could have at most  one trap at $f = 0$.

The control problem with objective functional ${\cal J}_A[f]$ for a two-level system can be conveniently described by the vectors
\begin{eqnarray*}
  {\bbr} &=& {\rm Tr}(\rho_0\boldsymbol{\sigma}), \\
  \ba &=& {\rm Tr}(e^{i\sigma_zT}Ae^{-i\sigma_zT}\boldsymbol{\sigma}), \\
  \bv &=& \frac{1}{2}{\rm Tr}(V\boldsymbol{\sigma}), \\
  {\bf h}_0&=&\Tr(\boldsymbol{\sigma}H_0)
\end{eqnarray*}
which define the initial state of the system, the target observable, the interaction potential and the free Hamiltonian, respectively.
Here $\bbr,\ba,\bv,{\bf h}_0\in \mathbb R^3$,  $|\bbr|\leq 1$ and  $\boldsymbol{\sigma}=(\sigma_x,\sigma_y,\sigma_z)$ is the vector of the Pauli matrices
\begin{equation}
\sigma_x=\Bigg(\begin{array}{*{20}{c}}0 & 1 \\ 1 & 0\end{array}\Bigg), \qquad  \sigma_y=\Bigg(\begin{array}{*{20}{c}}0 & -i \\ i & 0\end{array}\Bigg), \qquad \sigma_z=\Bigg(\begin{array}{*{20}{c}} 1 & 0 \\ 0 & -1\end{array}\Bigg).
\end{equation}
Any nontrivial (i.e., non-zero and not equal to identity) observable $A$ for a two-level system is canonically equivalent to a rank one projector and hence it is enough to consider target observables $A$ which are projectors, so that $|\ba|=1$. We can choose one of these vectors arbitrarily without loss of generality, so we set  ${\bf h}_0={\bf e}_z$. 

Let us show, for example, that any constant control $f\ne f_0$ can not be trap and any piecewise constant control composed of two pieces also can not be trap. Variations of the unitary evolution produced by variations $\delta f$ of the control have the following form:
\[
U_{T}^{f_0+\delta f}=U_{T}^{f_0}e^{-i\int_{0}^{T}V_t^{f_0}\delta f(t)dt}, \quad V_t^{f_0}=U_{t}^{f_0 \dag }VU_{t}^{f_0}
\]
In order to produce arbitrary variations in the neighborhood of the point $U_{T}^{f_0}$, the matrices $V_t^{f_0}$ should span the space of all $2\times2$ traceless Hermitian matrices $su(2)$. As an example, consider Landau-Zener system with Hamiltonian $H=\sigma_z+f\sigma_x$ and constant control pulse $f=const$.  Then $V_t^{f}=e^{i(\sigma_z+f\sigma_x)t}\sigma_x e^{-i(\sigma_z+f\sigma_x)t}=(R_{{\bf n},\omega t}{\bf e}_x)\cdot\boldsymbol{\sigma}$, where $R_{{\bf n},\omega t}{\bf e}_x$ is the result of rotation of the vector ${\bf e}_x$ around the axis ${\bf n}={\bf e}_z+f{\bf e}_x$ clockwise by the angle $\omega t$, $\omega=2(1+f^2)$. The matrices $V_t^{f}$ span $su(2)$ if the vector ${\bf n}$ is not orthogonal to the vector ${\bf e}_x$, otherwise, if ${\bf n}\perp{\bf e}_x$ the matrices $V_t^{f}$ span only the subspace of matrices of the form $\alpha\sigma_x+\beta\sigma_y$. Therefore, if ${\bf n}\cdot{\bf e}_x=f\neq0$ then $f$ is not a trap. Notice, that for Landau-Zener system $f_0=0$. Let us consider control composed of two constant pieces $g=f_1\theta(T/2-t)+f_2\theta(t-T/2)$, where $f_1, f_2=const$, $f_1\neq f_2$ and $\theta(t)$ is the Heaviside step function. If $f_1\neq f_2$ then the control $g$ is not a trap because the axes ${\bf n}_1={\bf e}_z+f_1{\bf e}_x$ and ${\bf n}_2={\bf e}_z+f_2{\bf e}_x$ are not parallel. Indeed, consider the set of vectors $S_1=\{{\bf x}:{\bf x}=R_{{\bf n_1},\omega_1 t}{\bf e}_x, 0\leq t \leq T/2\}$. This set can span ${\mathbb R}^3$ or at least the plane which is orthogonal to the vector ${\bf n}_1$. Similarly the set of vectors $S_2=\{{\bf x}:{\bf x}=R_{{\bf n_2},\omega_2 t}R_{{\bf n_1},\omega_1 T/2}{\bf e}_x, T/2\leq t \leq T\}$ span ${\mathbb R}^3$ or at least the plane which is orthogonal to the vector ${\bf n}_2$.   Since  vectors ${\bf n}_1$ and ${\bf n}_2$ are not parallel, the set $S=S_1\bigcup S_2$ span ${\mathbb R}^3$ anyway. The set $S_1$ corresponds to the matrices $V_t^{g}$, $0\leq t \leq T/2$ and the set $S_2$ corresponds to the matrices $V_t^{g}$, $T/2\leq t \leq T$. Thus, all together these matrices span $su(2)$. 

Next theorem states that traps do not exist if at least one of the vectors $\bbr$, $\ba$, or $\bv$ has non-zero $z$-th component~\cite{Mian2}.

\begin{theorem}\label{thZ}
For a system with evolution~(\ref{e01}) traps for $\J_A$ may exist only if $\bbr_z=\ba_z=\bv_z=0$. If at least one of the vectors $\bbr$, $\ba$, or $\bv$ has non-zero $z$-th component then $\J_A$ is trap-free for any $T$.
\end{theorem}

Theorem~\ref{thZ} shows that $f=f_0$ can be trap for small $T$ only if the vectors $\bbr$, $\ba$ and $\bv$  belong to the plane orthogonal to the vector ${\bf h}_0$.  In our case ${\bf h}_0={\bf e}_z$ and if at least one of the vectors $\bbr$, $\ba$ and $\bv$ has non-zero $z$-component, then the corresponding control problem has no traps for any arbitrarily small $T$. Therefore in the rest of this work we will analyze the case when $\bbr_z=\ba_z=\bv_z=0$. For physical interpretation of the conditions in Theorem~\ref{thZ} consider spin $1/2$ particle which interacts with constant magnetic field $B_z$ directed along the axis $Oz$ and with control magnetic field ${\bf B}={\bf n}f(t)$ directed along some vector $\bf n$. Then the Hamiltonian of the spin has the form $H(t)=\sigma_z B_z+(\boldsymbol{\sigma}, {\bf n}) f(t)$. As the control goal, consider preparation of some target spin state $|\psi_{\rm f}\rangle$ so that $A=|\psi_{\rm f} \rangle \langle\psi_{\rm f}|$.  Then the condition $\bbr_z=\ba_z=0$  means that projection of spin on z-axis in the initial and final states should be zero. The condition $\bv_z=0$ means that the control magnetic field should be orthogonal to $B_z$.

Define $\bbr_{0}=\sin\phi{\bf e}_x+\cos\phi{\bf e}_y=({\bf e}_z \times {\bv})/{v}$, $\bbr_{k}=\sin(2t_k-\phi){\bf e}_x+\cos(2t_k-\phi){\bf e}_y,~k=1,2$, where $\phi=\arctan(v_y/v_x)$, $\bbr_{k}=\sin(2t_k-\phi){\bf e}_x+\cos(2t_k-\phi){\bf e}_y,~k=1,2$  and $v=|\bv|$. Denote $\p=\p(\bbr,\ba, \bv)=v^2(\bbr\cdot\bbr_{0})(\ba\cdot\bbr_{0})=(\bv\times\bbr)_z(\bv\times\ba)_z$ and $\q=\q(\bbr,\ba, \bv)=(\bbr\times \ba)_z$.

Consider four domains in the parameter space:

Domain $\D_I=\{(\bbr,\ba, \bv)\,|\, \bbr_z=\ba_z=\bv_z=0, \p>0, \q>0\}$.

Domain $\D_{II}=\{(\bbr,\ba, \bv)\,|\, \bbr_z=\ba_z=\bv_z=0, \p<0, \q<0\}$.

Domain $\D_{III}=\{(\bbr,\ba, \bv)\,|\, \bbr_z=\ba_z=\bv_z=0, \p>0, \q<0\}$.

Domain $\D_{IV}=\{(\bbr,\ba, \bv)\,|\, \bbr_z=\ba_z=\bv_z=0, \p<0, \q>0\}$.

\begin{figure}[h!]\center
\includegraphics[width=.7\linewidth]{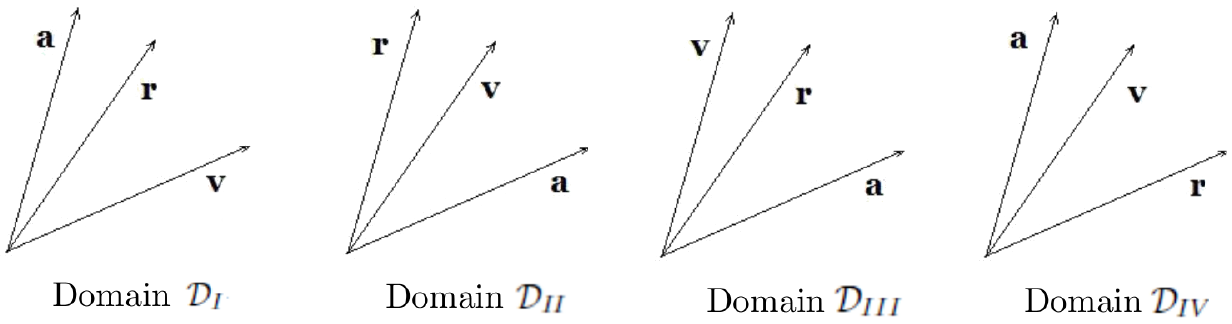}
\caption{The mutual arrangement of the vectors $\bf r$, $\bf a$ and $\bf v$ in the domains $\D_{I}$, $\D_{II}$, $\D_{III}$ and  $\D_{IV}$.\label{fig1}}
\end{figure}

The mutual arrangement of the vectors $\bf r$, $\bf a$ and $\bf v$ in the domains $\D_I$, $\D_{II}$, $\D_{III}$ and $\D_{IV}$ is shown in Fig.~\ref{fig1}.

\section{Absence of traps when the parameters belong to the domains $\D_{III}$ and $\D_{IV}$}

In this section we prove that if the parameters belong to the domain $\D_{III}$ or $\D_{IV}$, then the objective has no traps.  Our analysis  will use the expression for the Hessian ${\rm Hess}_{f}{\cal J}_A(\tau_2,\tau_1)$ of the objective functional $\J_A[f]$ at the point $f=0$:
\begin{eqnarray*}
&&{\rm Hess}_{f}{\cal J}_A(t_2,t_1)\bigg|_{f=0}=\frac{\delta^2 {\cal J}_A[f]}{\delta f(t_2)\delta f(t_1)}\bigg|_{f=0}=-{\rm Tr}([V_{t_1} ,[V_{t_2} , \rho_0]]A_T)\bigg|_{f=0} \\
&=&-{v^{2}}\bigg[r_x \sin(2t_2-\phi)+r_y \cos(2t_2-\phi)\bigg]\bigg[a_x \sin(2t_1-\phi)+a_y \cos(2t_1-\phi)\bigg],~~~ t_1> t_2.
\end{eqnarray*}
Here $A_T:=e^{i\sigma_zT}Ae^{-i\sigma_zT}$. Expressing the Hessian in terms of the vectors $\bbr_{k}=\sin(2t_k-\phi){\bf e}_x+\cos(2t_k-\phi){\bf e}_y,~k=1,2$, we obtain
\begin{equation}\label{Hess}
{\rm Hess}_{f}{\cal J}_A(t_2,t_1)= \left\{
\begin{array}{l}
-{v^{2}}(\bbr\cdot\bbr_{2})(\ba\cdot\bbr_{1}), \qquad t_1\geqslant t_2 \vspace{1mm} \\
-{v^{2}}(\bbr\cdot\bbr_{1})(\ba\cdot\bbr_{2}), \qquad t_1< t_2
\end{array}\right.
\end{equation}
This expression will play the key role in the analysis below.

\begin{theorem}\label{saddleth}
If $(\bbr,\ba,\bv)$ belong to the domain  $\D_{III}$ or $\D_{IV}$, then the control $f=0$ is a saddle point for any $T>0$.
\end{theorem}
\noindent{\bf Proof.}
Let $\delta_{\varepsilon} (t)$ be an approximation of Dirac delta function:
\begin{equation}
\delta_{\varepsilon} (t)=\left\{
\begin{array}{l}
0, \qquad  |t|\geqslant \frac{\varepsilon}{2} \\
\frac{1}{\varepsilon}, \qquad |t|< \frac{\varepsilon}{2}
\end{array}\right.
\end{equation}
We have $\forall f\in {\rm C}[0,T]$, $\forall t\in(\varepsilon / 2,~T-\varepsilon / 2)$, where $\varepsilon<T$
\begin{equation}
\int\limits_0^T \delta_{\varepsilon} (\tau-t) f(\tau)d\tau=f(t)+O(\varepsilon)
\end{equation}
Let $f_{\varepsilon}(t)=\lambda \delta_{\varepsilon} (t-t_1)+\mu \delta_{\varepsilon} (t-t_2)$,~$\varepsilon / 2<t_1<t_2<T-\varepsilon / 2$,~$\varepsilon<t_2-t_1$. Then, substituting the function  $f_{\varepsilon}(t)$ in the expression for the Hessian ($\ref{Hess}$), we obtain
\begin{equation}\label{e2}
(f_{\varepsilon},Hf_{\varepsilon})=-{v^{2}}\bigg[  \lambda^{2}(\bbr\cdot\bbr_{2})(\ba\cdot\bbr_{2})+2\lambda\mu(\bbr\cdot\bbr_1)(\ba\cdot\bbr_2)+\mu^{2}(\bbr\cdot\bbr_1)(\ba\cdot\bbr_1)\bigg]   +O(\varepsilon)
\end{equation}
Denote the quadratic form in the variables  $\lambda, \mu$ in the square brackets in eq.~(\ref{e2}) as
\begin{equation}
G(\lambda,\mu)=\lambda^{2}(\bbr\cdot\bbr_{2})(\ba\cdot\bbr_{2})+2\lambda\mu(\bbr\cdot\bbr_1)(\ba\cdot\bbr_2)+\mu^{2}(\bbr\cdot\bbr_1)(\ba\cdot\bbr_1)
\end{equation}
Quadratic form $G(\lambda,\mu)$  takes positive and negative values if and only if it has  positive discriminant, i.e. if and only if
\begin{equation}\label{e22}
(\bbr\cdot\bbr_1)^2(\ba\cdot\bbr_2)^2-(\bbr\cdot\bbr_2)(\ba\cdot\bbr_2)(\bbr\cdot\bbr_1)(\ba\cdot\bbr_1)>0
\end{equation}
The quantity $(\bbr\cdot\bbr_1)(\ba\cdot\bbr_2)$ is a continuous function of the arguments $t_1$ and $t_2$ and therefore in some neighbourhood of  $t_1=t_2=0$ the asymptotic expression $(\bbr\cdot\bbr_1)(\ba\cdot\bbr_2)=(\bbr\cdot\bbr_0)(\ba\cdot\bbr_0)+O(T)$ is satisfied. Hence there exist $t_1$ and $t_2$, $0<t_1<t_2<T$ such that in the domain $\D_{III}$ the inequality   $(\bbr\cdot\bbr_1)(\ba\cdot\bbr_2)>0$ holds. The inequality~(\ref{e22})  then implies that
\begin{equation}\label{e3}
(\bbr\cdot\bbr_1)(\ba\cdot\bbr_2)-(\bbr\cdot\bbr_2)(\ba\cdot\bbr_1)>0
\end{equation}
Using the identity $(\ba\times{\bf b})\cdot({\bf c}\times{\bf d})=(\ba\cdot{\bf c})({\bf b}\cdot{\bf d})-(\ba\cdot{\bf d})({\bf b}\cdot{\bf c})$, the inequality (\ref{e3}) can be transformed to the inequality
\begin{equation}\label{e4}
(\bbr\times\ba)\cdot(\bbr_1\times\bbr_2)>0
\end{equation}
Since  $\bbr_1\times\bbr_2=-\sin 2(t_2-t_1) {\bf e}_z$,  the inequality (\ref{e4}) for $t_2>t_1$  implies the inequality $(\bbr\times\ba)_z<0$.
Hence, we obtain that in the domain $\D_{III}$ the quadratic form  $G(\lambda,\mu)$ takes positive and negative values. Let us choose $\lambda_1, \mu_1$ such that $G(\lambda_1,\mu_1)>0$ and $\lambda_2, \mu_2$ such that $G(\lambda_2,\mu_2)<0$ and consider
\begin{eqnarray*}
f_{\tilde{\varepsilon}, 1 }(t)&=&\lambda_1 \delta_{\tilde{\varepsilon}} (t-t_1)+\mu_1 \delta_{\tilde{\varepsilon}} (t-t_2)\\
f_{\tilde{\varepsilon}, 2 }(t)&=&\lambda_2 \delta_{\tilde{\varepsilon}} (t-t_1)+\mu_2 \delta_{\tilde{\varepsilon}} (t-t_2)
\end{eqnarray*}
where $\tilde{\varepsilon}$ is small enough such that $(f_{\tilde{\varepsilon},j},Hf_{\tilde{\varepsilon},j})$ for $j=1,2$ have the same sign as  $\lim\limits_{\varepsilon\rightarrow 0}(f_{\varepsilon,j},Hf_{\varepsilon,j})$. Then $(f_{\tilde{\varepsilon},1},Hf_{\tilde{\varepsilon},1})<0$ and  $(f_{\tilde{\varepsilon},2},Hf_{\tilde{\varepsilon},2})>0$ and hence Hessian at $f=0$ has positive and negative eigenvalues so that $f=0$ is a saddle.

Similarly the statement of the theorem can be proved for the domain $\D_{IV}$. Indeed, if we consider the domain $\D_{IV}$, then  there exist $t_1$ and $t_2$, $0<t_1<t_2<T$,  such that  the inequality   $(\bbr\cdot\bbr_1)(\ba\cdot\bbr_2)<0$ holds and the inequality~(\ref{e22}) implies the inequality
\begin{equation}\label{e31}
(\bbr\cdot\bbr_1)(\ba\cdot\bbr_2)-(\bbr\cdot\bbr_2)(\ba\cdot\bbr_1)<0
\end{equation}
which  is equivalent to the inequality
\begin{equation}\label{e41}
(\bbr\times\ba)\cdot(\bbr_1\times\bbr_2)<0
\end{equation}
Hence, similar to the case of the domain $\D_{III}$, we conclude that $(\bbr\times\ba)_z>0$ and that in the domain  $\D_{IV}$ the quadratic form $G(\lambda,\mu)$ also takes positive and negative values. Choose again $\lambda_1, \mu_1$ and $\lambda_2, \mu_2$ such that $G(\lambda_1,\mu_1)>0$ and $G(\lambda_2,\mu_2)<0$ and consider
\begin{eqnarray*}
f_{\tilde{\varepsilon}, 1 }(t)&=&\lambda_1 \delta_{\tilde{\varepsilon}} (t-t_1)+\mu_1 \delta_{\tilde{\varepsilon}} (t-t_2)\\
f_{\tilde{\varepsilon}, 2 }(t)&=&\lambda_2 \delta_{\tilde{\varepsilon}} (t-t_1)+\mu_2 \delta_{\tilde{\varepsilon}} (t-t_2)
\end{eqnarray*}
where $\tilde{\varepsilon}$ is such that $(f_{\tilde{\varepsilon},j},Hf_{\tilde{\varepsilon},j})$ for $j=1,2$ have the same sign as $\lim\limits_{\varepsilon\rightarrow 0}(f_{\varepsilon,j},Hf_{\varepsilon,j})$.

Thus, for the domains $\D_{III}$ and $\D_{IV}$ for all  $T>0$ there exist functions $f_1=f_{\tilde{\varepsilon}, 1 }$ and $f_2=f_{\tilde{\varepsilon}, 2 }$ such that $(f_1,Hf_1)<0$ and $(f_2,Hf_2)>0$. Therefore Hessian at $f=0$ has positive and negative eigenvalues so that $f=0$ is a saddle. This completes the proof.

Theorem \ref{saddleth} implies that if the parameters of the control problem belong to the domains $\D_{III}$ and $\D_{IV}$, then the control problem is trap-free for any $T$. Below we express the condition for the absence of traps instead of the vector $\ba$ in terms of the vector $\ba_0 = {\rm Tr}(A\boldsymbol{\sigma})$. This condition can be practically more convenient since vector $\ba_0$ directly determines the target observable.

\begin{theorem} If the parameters $\bbr$, $\ba_0$, $\bv$, and $T $ satisfy the relation
\begin{equation}\label{e7}
(\bv\times\bbr)_z[(\bv\times\ba_0)_z\cos 2T-(\bv\cdot\ba_0)\sin 2T][(\bbr\times\ba_0)_z\cos 2T-(\bbr\cdot\ba_0)\sin 2T]<0.
\end{equation}
then the control objective has no traps. If the parameters $\bbr$, $\ba_0$, and $\bv$ satisfy the relation
\begin{equation}\label{e9}
(\bv\times\bbr)_z(\bv\times\ba_0)_z(\bbr\times \ba_0)_z<0
\end{equation}
then there exists $\tilde T$ such that for all $T<\tilde T$ the control objective has no traps.
\end{theorem}
\noindent{\bf Proof.} The condition which defines the domains $\D_{III}$ and $\D_{IV}$ and hence implies the absence of traps can be rewritten as
\begin{equation}\label{e5}
\p(\bbr,\ba, \bv)\q(\bbr,\ba, \bv)=(\bv\times\bbr)_z(\bv\times\ba)_z(\bbr\times \ba)_z<0
\end{equation}
Since
\begin{equation}\label{e6}
\ba=\ba_0\cos 2T+\ba_0\times{\bf e}_z\sin 2T,
\end{equation}
the inequality~(\ref{e5}) can be transformed into the inequality
\[
(\bv\times\bbr)_z[(\bv\times\ba_0)_z\cos 2T-(\bv\cdot\ba_0)\sin 2T][(\bbr\times\ba_0)_z\cos 2T-(\bbr\cdot\ba_0)\sin 2T]<0.
\]
This inequality can be expressed in terms of the angles $\alpha=\angle(\bv,\ba_0)$ and $\beta=\angle(\bbr,\ba_0)$  as
\begin{equation}\label{e8}
\sin (\alpha-\beta)\sin (2T-\alpha)\sin (2T-\beta)<0.
\end{equation}
Note that $-\pi<\alpha<\pi$, $-\pi<\beta<\pi$, $\alpha\neq 0$, $\beta\neq 0$ and $\alpha-\beta=\angle(\bv,\bbr)$. Therefore for any $\bbr$, $\ba_0$, and $\bv$ such that
\[
(\bv\times\bbr)_z(\bv\times\ba_0)_z(\bbr\times \ba_0)_z<0
\]
or, equivalently, for $\alpha$ and $\beta$ such that
\begin{equation}\label{e10}
\sin (\alpha-\beta)\sin \alpha\sin \beta<0
\end{equation}
there exists $\tilde T$ such that for all $T<\tilde T$ the inequality (\ref{e10}) implies the inequality (\ref{e8}) and, hence, the control objective has no traps. This finished the proof of the Theorem.

\section{Example: absence of traps for spin rotation}
In this section we consider as an example manipulation by direction of spin with ultrafast control. Let $|\hspace{-.1cm}\uparrow \rangle$ and $|\hspace{-.1cm}\downarrow \rangle$ be states with direction of spin up and down, respectively. The control problem which we consider is to maximize the transition probability  from the initial state $(i|\hspace{-.1cm}\uparrow \rangle+|\hspace{-.1cm}\downarrow \rangle)/\sqrt{2}$ with spin  directed at the initial time $t=0$ along the negative direction of the axis $0y$ to the target state $(|\hspace{-.1cm}\uparrow \rangle + |\hspace{-.1cm}\downarrow \rangle)/\sqrt{2}$ with spin  directed at the final time $T$ along the positive direction of the axis $Ox$.
\begin{figure}[h!]\center
\includegraphics[width=.5\linewidth]{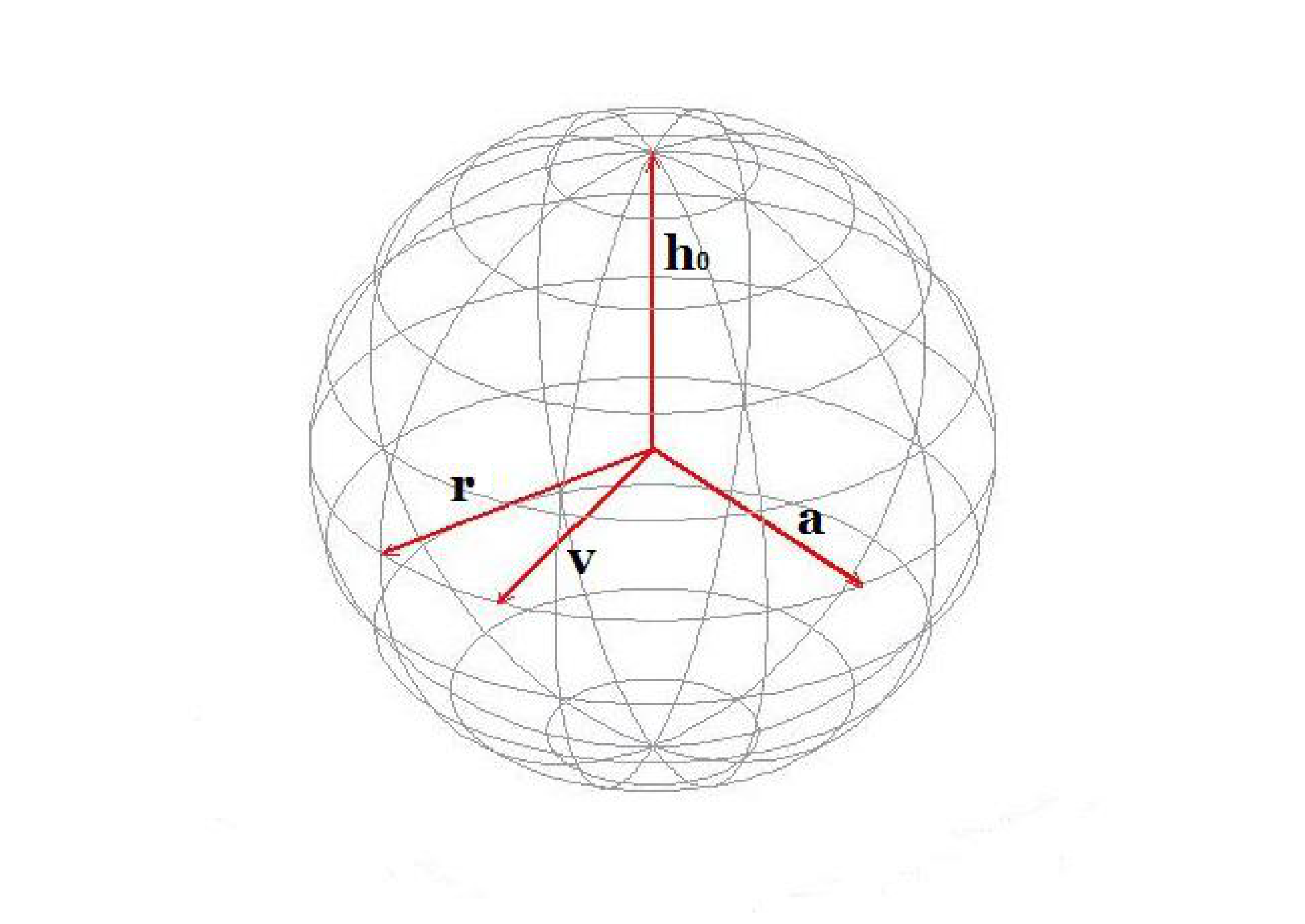}
\caption{Vectors $\bbr$, $\ba$, $\bv$ and ${\bf h_0}$ for  rotation of spin from the state $(i|\hspace{-.1cm}\uparrow \rangle+|\hspace{-.1cm}\downarrow \rangle)/\sqrt{2}$ to the state $(|\hspace{-.1cm}\uparrow \rangle + |\hspace{-.1cm}\downarrow \rangle)/\sqrt{2}$.\label{fig4}}
\end{figure}

The initial density matrix of the spin for this problem is
\begin{equation}
\rho_0=\frac{1}{2}\left(1-\sigma_y\right)
\end{equation}
The objective functional ${\cal J}_A[f] = {\rm Tr}(\rho_T A)$ is determined by the matrix
\begin{equation}
A=\frac{1}{2}\left(1+\sigma_x\right),
\end{equation}
For this problem the parameters have the form $\bbr=-{\bf e}_y$, $\ba=\cos 2T{\bf e}_x-\sin 2T {\bf e}_y$, and $\bv=v_x{\bf e}_x+v_y{\bf e}_y$. The  vectors $\bbr$, $\ba$, $\bv$ and ${\bf h_0}$ are shown on Fig.~\ref{fig4}.
These parameters    belong to the domain $\D_{III}\bigcup\D_{IV}$ if
\begin{equation}\label{Ineq1}
(\bv\times\bbr)_z(\bv\times\ba)_z(\bbr\times \ba)_z=v_x(v_x\sin 2T +v_y\cos 2T)\cos 2T<0.
\end{equation}
The inequality (\ref{Ineq1})  can be rewritten as
\begin{equation}\label{tg1}
\tan 2T<-\frac{v_y}{v_x}, \quad \cos 2T\ne 0, \quad v_x\ne 0.
\end{equation}
According to Theorem~\ref{saddleth}, the exceptional control $f_0$ is a saddle  of the objective functional $\J_A[f]$ for any $T$ satisfying~(\ref{tg1}).

\section{Numerical analysis of the control landscape}

In this section we perform numerical analysis of the control landscape in order to illustrate the statement of the Theorem~\ref{saddleth}.

\begin{figure}[h!]\center
\includegraphics[width=1\linewidth]{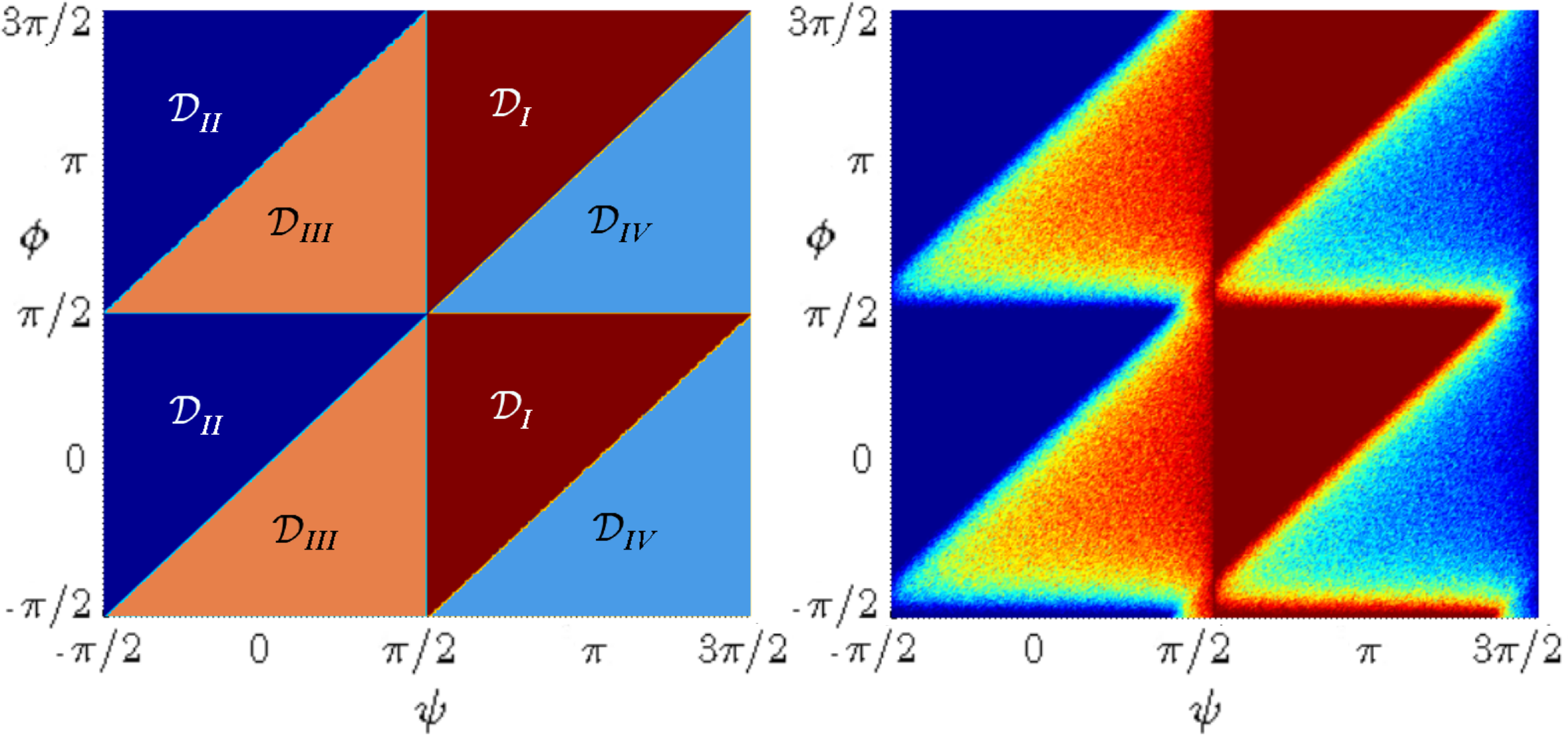}
\caption{Domains $\D_{I}$, $\D_{II}$, $\D_{III}$, $\D_{IV}$ (left subplot) and numerically estimated probability $P$ of the inequality $\J[f]<\J_A[0]$ for  $T=\pi/12$   (right subplot). Red color corresponds to $P=1$ and blue color to $P=0$. Probability $P$ for each pair $(\phi,\psi)$ is computed as fraction of the  number of realizations $N_{\J[f]<\J_A[0]}$  of the inequality $\J[f]<\J_A[0]$ among $300$ values of $\J[f]$. Controls are piecewise constant functions $f=\sum_{i=1}^{100} a_i\chi_i$, where $\chi_i$ is the characteristic function of the interval $[(i-1)T/100, iT/100]$ and each $a_i$ has  normal distribution with unit dispersion. For the domains $\D_{III}$ and $\D_{IV}$ the probability $P$ should take intermediate values between zero and one. We see that the domains $\D_{III}$ and $\D_{IV}$ on the left subplot are in good agreement with the area on the right subplot where $0<P<1$. \label{fig5}}
\end{figure}

We choose vector $\bbr$ of the initial density matrix $\rho_0$ to be $\bbr={\bf e}_y$ and consider vectors $\bv$ and $\ba$ to be of unit norm and belonging to the plane $Oxy$. In this case $\bv=\cos\phi {\bf e}_x+\sin\phi {\bf e}_y$ and $\ba=\cos\psi {\bf e}_x+\sin\psi {\bf e}_y$, where $\phi$ and $\psi$ are  the angles between vectors $\bv$, $\ba$ and $Ox$ axis. The control problem is parametrized by the angles $\phi$ and $\psi$. In this case we can restrict the domains $\D_{I}$, $\D_{II}$, $\D_{III}$, and $\D_{IV}$ to the variables $\phi$ and $\psi$. These restricted domains are shown on the left subplot of Fig.~\ref{fig5}.

\begin{figure}[h!]\center
\includegraphics[width=1\linewidth]{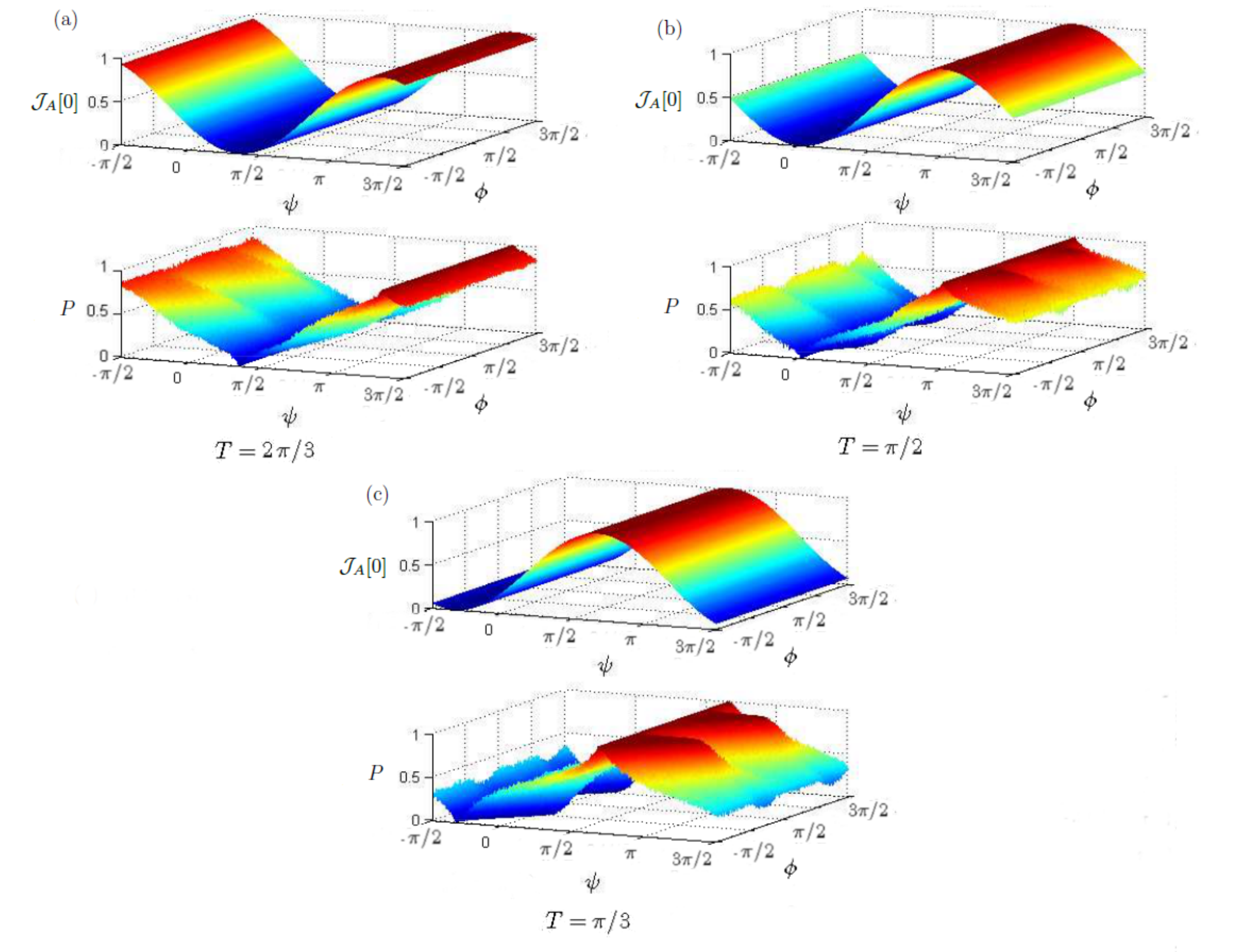}
\caption{Graphs of $\J_A[0]$ and of probability $P$ that $\J_A[f]<\J_A[0]$ vs. angles $\phi$ and $\psi$ for (a)  $T=2\pi/3$, (b)  $T=\pi/2$ and (c)  $T=\pi/3$. Probability $P$  for each pair $(\phi,\psi)$ is the fraction of number of realizations $N_{\J_A[f]<\J_A[0]}$  of the inequality $\J_A[f]<\J_A[0]$ among $300$ values of $\J_A[f]$. Controls are piecewise constant functions $f=\sum_{i=1}^{100} a_i\chi_i$, where $\chi_i$ is the characteristic function of the interval $[(i-1)T/100, iT/100]$ and each $a_i$ has normal distribution with unit dispersion. For  $T=2\pi/3>\pi/2$ probability $P$ reaches its maximum only if the angles $(\psi,\phi)$ are such that $\J_A[0]=1$; in this case the exceptional control $f=0$ is a global maximum but not a trap. \label{fig6}}
\end{figure}

For each pair $(\phi,\psi)$ we randomly select $300$ controls $f$ and evaluate the corresponding objective values $\J_A[f]$ for $T=\pi/12$. Controls are chosen as piecewise constant functions $f=\sum_{i=1}^{100} a_i\chi_i$, where $\chi_i$ is the characteristic function of the interval $[(i-1)T/100, iT/100]$ and each $a_i$ has a normal distribution centered at zero and with unit dispersion.  Then we compare the value $\J_A[0]$ of the objective $\J_A[f]$ at the point $f=0$ with its values in some neighborhood of this point. Left subplot shows numerically estimated probability $P(\phi,\psi)$ of the inequality $\J_A[f]<\J_A[0]$  which is computed as fraction of the number of realizations $N_{\J_A[f]<\J_A[0]}$  of the inequality $\J_A[f]<\J_A[0]$ among $300$ values of $\J_A[f]$ for each pair $(\phi,\psi)$,
\begin{equation}
P(\phi,\psi)=\frac{N_{\J_A[f]<\J_A[0]}}{300}.
\end{equation}
According  to Theorem~\ref{saddleth}, if the parameters of the control problem belong to the domain $\D_{III}$ or to the domain $\D_{IV}$, in the neighborhood of $f=0$ some functions should satisfy $\J_A[f]<\J_A[0]$ while other functions should satisfy $\J_A[f]>\J_A[0]$. Hence for $\phi$ and $\psi$ such that $(\bbr,\ba,\bv)$ belong to the domain $\D_{III}$ or to the domain $\D_{IV}$, the probability $P$ should satisfy $0<P<1$. This statement agrees with the numerical results shown in right subplot of Fig.~\ref{fig5}. Dark blue and dark red areas correspond to the cases $P=0$ and $P=1$, respectively, while areas with intermediate colors correspond to the case $0<P<1$, i.e., to saddle. Left subplot of Fig.~\ref{fig5} shows the theoretically computed domains $\D_{I}$, $\D_{II}$, $\D_{III}$, and $\D_{IV}$ in terms of the angles $\phi$ and $\psi$. We see that areas with intermediate colors on the right subplot are in good agreement with domains $\D_{III}$ and $\D_{IV}$ on the left subplot.

In~\cite{Mian3} it was proved that for $T>\pi/2$ the objective $\J_A[f]$ has no traps. Fig.~\ref{fig6} shows the graph of $\J_A[0]$ as function of the angles $(\psi,\phi)$ and the probability $P$ of the inequality $\J_A[f]<\J_A[0]$. We see that for $T=2\pi/3>\pi/2$ the probability $P$ reaches its maximum $P=1$ in the same area $\D_{\rm max}$ where $\J_A[0]$ reaches global maximum, $\J_A[0]=1$. Hence the exceptional control $f=0$ for $T>\pi/2$  is not a trap of $\J_A[f]$ for any $(\psi,\phi)$. For $T<\pi/2$ areas where $P=1$ and $J_A[0]=1$ do not coincide.

\newpage
\section{Conclusions}
In this paper we study ultrafast control of a qubit and analyze local extrema of transition probability in the qubit for a short duration $T$ of the control pulse.  Hamiltonian of the qubit is $H=\sigma_z+f(t)(v_x\sigma_x+v_y\sigma_y)$. Any control $f\neq 0$  is not a trap and, moreover, $f=0$ is not a trap if at least one of the vectors  $\bv={\rm Tr}(V\boldsymbol{\sigma})/2$, $\bbr={\rm Tr}(\rho_0\boldsymbol{\sigma})$, or $\ba_0={\rm Tr}(A\boldsymbol{\sigma})$ does not belong to the plane $Oxy$. We show that if $\bv$, $\bbr$, and $\ba_0$ belong to the plane  $Oxy$ then control $f=0$ is a saddle point if the parameters satisfy  the relation~(\ref{e7}). In this case the control problem is trap free. Moreover, for any $\bv$, $\bbr$, $\ba_0$ which satisfy~(\ref{e9}) there exists $\tilde T$ such that for all $T<\tilde T$ the control problem is trap free.

In the analysis of this work, we neglect the interaction of the system with the environment. Such approximation is reasonable if the duration of the control pulse is smaller than the typical decoherence time of the system. If decoherence is faster than control duration then the analysis of the control landscape in the presence of dissipative, e.g., Lindbladian dynamics, is needed. Another factor which may affect the structure of the control landscape is the noise in the control which can influence the landscape by decreasing the maximal objective value~\cite{Hocker2014}.  Our analysis uses essentially unitary dynamics and would requires further explorations for open or noisy quantum systems.

\end{document}